\documentclass[aps,pra,twocolumn,superscriptaddress,10pt]{revtex4-1}
\usepackage[utf8]{inputenc}
\usepackage{cochineal}
\usepackage[cochineal]{newtxmath}
\usepackage{mathtools,graphicx,microtype,bm,cases,siunitx}
\usepackage[cal=boondoxo,frak=boondox]{mathalfa}
\usepackage[colorlinks=true,allcolors=blue]{hyperref}

\begin{document}

\title{Bound electron pairs formed by the spin-orbit interaction in 2D gated structures}

\author{Yasha Gindikin}
\affiliation{Kotelnikov Institute of Radio Engineering and Electronics, Russian Academy of Sciences, Fryazino, 141190, Russia
}

\author{Vitalina Vigdorchik}
\affiliation{Faculty of Physics, Lomonosov Moscow State University, Moscow 119991, Russia}

\author{Vladimir A.\ Sablikov}
\affiliation{Kotelnikov Institute of Radio Engineering and Electronics, Russian Academy of Sciences, Fryazino, 141190, Russia
}

\begin{abstract}
We explore the bound electron pairs (BEPs) arising due to the pair spin-orbit interaction (PSOI) in two-dimensional structures with a gate that can allow the BEPs to be manipulated. The gate breaks the in-plane reflection symmetry of the pair Coulomb field and creates a one-particle Rashba spin-orbit interaction. We find that the normal component of the electric field substantially affects the BEPs but the key role in forming the BEPs belongs to the in-plane component. The ground state of a BEP with zero total momentum, which is doubly degenerate in the absence of the gate, splits into two states. One of them is tunable by varying the gate voltage whereas the other is on the contrary robust. The tunable BEP has a higher binding energy which grows as the gate voltage increases, with its orbital and spin structure changing continuously. At the large negative voltage the tunable BEP decays. The orbital and spin structure of the robust BEP does not depend on the gate voltage. Its energy level crosses the conduction band bottom at high gate voltage of any polarity, but the robust BEP remains bound and localized even when in continuum. 
\end{abstract}

\maketitle 

\section{Introduction}
The pair interaction of particles depends not only on the their charge and mutual distance but also on their spins and momenta. This well-known fact of the relativistic quantum mechanics~\cite{bethe2012quantum} is still too poorly studied for the electrons in crystals. However, it becomes important for modern materials with a strong Rashba spin-orbit interaction (SOI). 

In the relativistic quantum mechanics, the pair interaction of the electrons moving with small velocity $v/c\ll 1$ is described by the Breit-Pauli Hamiltonian~\cite{bethe2012quantum}. The pair interaction Hamiltonian derived from it in the frame of the $k \cdot p$ approximation~\cite{voon2009kp} has a form very similar to the original Breit-Pauli Hamiltonian with an important difference that the material-dependent coefficients appear in each of its terms. 

Of most interest is the SOI component of the pair interaction because it couples the spin and orbital degrees of freedom, which can essentially affect the dynamics of interacting electrons and result in new collective states. The pair spin-orbit interaction (PSOI) produced by the Coulomb fields $\mathbf{E}(\mathbf{r}_{i} - \mathbf{r}_{j})$ of interacting electrons has the following form~\cite{2019arXiv190506340G}
\begin{equation}
	\label{pse}
	H_{\mathrm{PSOI}} = \frac{\alpha}{\hbar} \sum_{i \ne j} \left( \mathbf{p}_{i} \times \mathbf{E} (\mathbf{r}_{i} - \mathbf{r}_{j})  \right) \cdot \bm{\sigma}_{i} \,,
\end{equation}
where $\mathbf{p}_{i}$ is the momentum of the $i$-th electron, $\bm{\sigma}$ is the Pauli vector, and $\alpha$ is a material-dependent SOI constant. Having been calculated within the $k \cdot p$ approximation, when the Coulomb field is assumed to be a smooth function on the scale of the lattice constant, the value of $\alpha$ is the same as the Rashba constant of the material.

The most interesting feature of the PSOI is that it creates attraction between the electrons in certain spin configurations tied to their momenta. A completely unusual property of such an attraction is the fact that it is determined directly by the electric field and, therefore, for the Coulomb interaction it is especially large at small distances between the particles, and rapidly decreases at large distances. Thus, the pair interaction we consider here is attractive on a small scale and repulsive on the large distance. The bound electron pairs (BEPs) formed as a result of this interaction~\cite{2018arXiv180410826G,PhysRevB.98.115137,GINDIKIN2019187} are drastically different from other composite particles, which are currently widely studied in bulk materials~\cite{combescot2015excitons,Kezerashvili2019}, low-dimensional systems~\cite{PhysRevB.81.045428,PhysRevB.92.085409,doi:10.1002/pssb.201800584,PhysRevB.95.085417,Mahajan_2006}, and even for cold atoms in optical lattice~\cite{winkler}.

Another feature of the PSOI is that it depends on the configuration of the Coulomb fields which act between electrons and can be controlled by a gate in low-dimensional systems. In order to find out how the field configuration affects BEPs and, in particular, their binding energy characterizing the stability of the pairs, we focus here on considering the two-body problem, which can be solved exactly. Bearing in mind possible implementations, the study of isolated pairs is of interest for low-dimensional structures, such as quantum dots and quantum cavities. The properties of isolated pairs are also worth studying for two-dimensional materials with not very high Fermi energy, when the Fermi wave vector is small on the scale of the reciprocal value of the region size, where the pair attraction prevails. In this case, the electrons near and below the Fermi energy as well as the pairs strongly repel each other, and therefore the Cooper instability does not occur.

In a two-dimensional (2D) system symmetric at the in-plane reflection, the PSOI is created by the electric field acting in the plane of the system~\cite{PhysRevB.98.115137,GINDIKIN2019187}. On the contrary, in one-dimensional (1D) systems the PSOI is created only by a normal component of the field, which arises when the axial symmetry of the system is broken e.g.\ by a proximate gate. In this case the PSOI originates from image charges induced by the interacting electrons at the gate~\cite{PhysRevB.95.045138}, and also leads to the BEP formation~\cite{2018arXiv180410826G}. 

The spectrum of the BEPs together with their spin structure are quite different depending on whether the PSOI is produced solely by the in-plane field or by the normal electric field. In spite of the difference in the electric field configuration in both cases, one can still classify the BEPs according to the nature of the electron motion, which produces the PSOI, to arrive at two distinct types of BEPs. 

The relative motion of electrons in the pair creates the \emph{relative} BEPs. In the symmetric 2D case the relative BEPs are triplet-like states with parallel spins, with the ground state of the BEP being doubly degenerate. In contrast, the relative BEPs in the 1D gated wire are of a mixed singlet-triplet type. 

The motion of an electron pair as a whole forms the \emph{convective} BEPs. Their binding energy crucially depends on the total momentum of the pair and the spin structure is more complicated. In symmetric 2D systems the convective BEPs do not possess a definite spin. Quite the opposite, in the gated 1D wire the spin projection is well defined so that $S_z = \pm 1$, its sign being determined by the direction of the center-of-mass momentum.

\begin{figure}[htb]
	\includegraphics[width=0.9\linewidth]{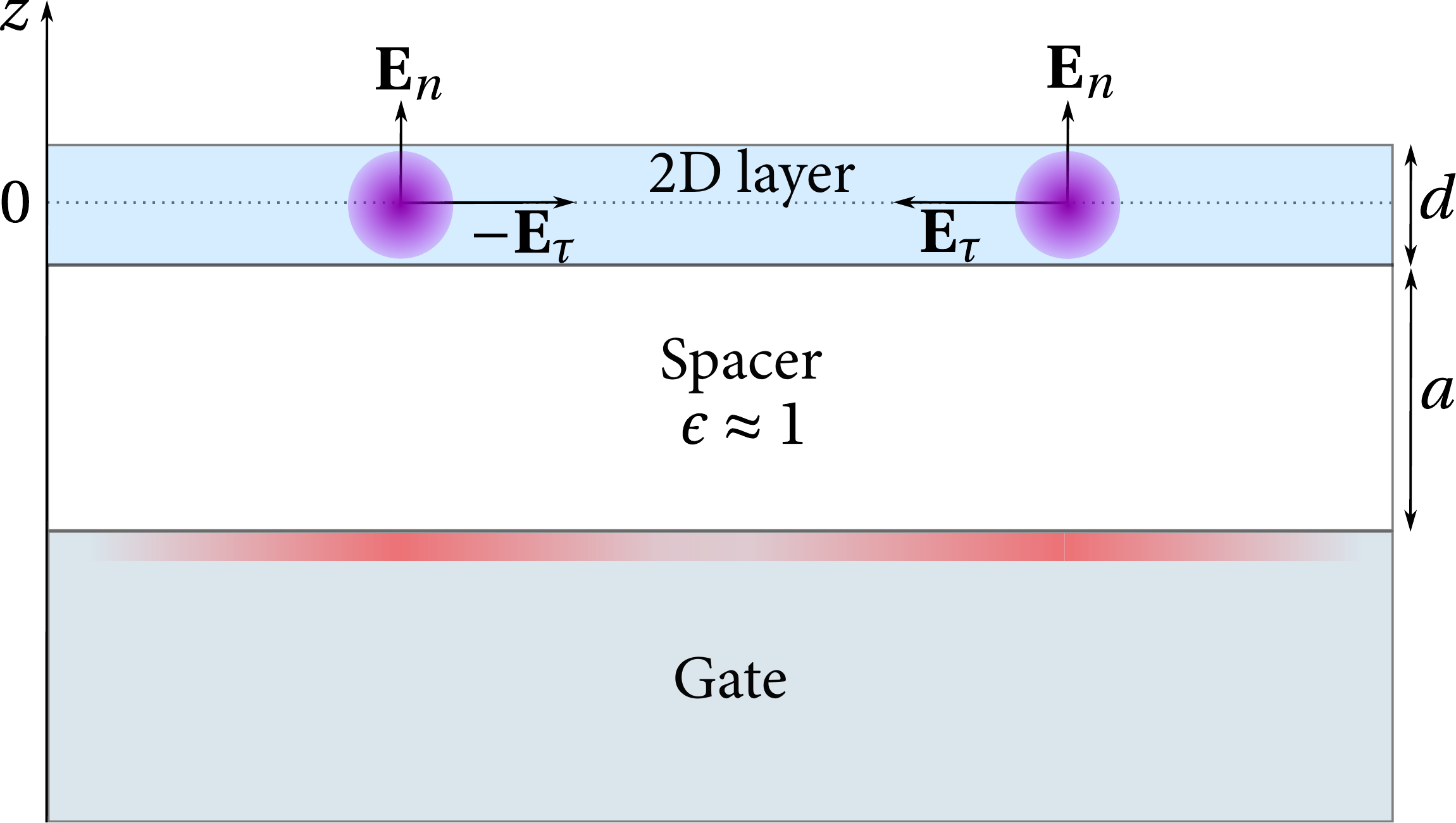}
		\caption{A 2D layer separated from a metallic gate by a spacer made of a weak dielectric. Normal ($\mathbf{E}_{n}$) and in-plane ($\mathbf{E}_{\tau}$) components of the electric field acting on each electron are created by a neighboring electron and the polarization charges, as well as the total charge of the gate.
	\label{fig1}}
\end{figure}

Of great interest is the problem of BEPs in 2D systems with a metal gate, since one can anticipate a unique opportunity to control their binding energy and the spin state. This raises the question of how the properties of BEPs change in such structures. The main effects are due to the fact that the symmetry of the Coulomb fields inherent in the two cases mentioned above is broken in such structures. The electric field has both in-plane and normal components, the interplay of which creates a nontrivial configuration of the effective magnetic field acting on the electron spin. The BEP spin state is not predefined, but rather should be determined self-consistently together with its orbital structure via the quantum-mechanical equations of motion taking into account the particular field configuration. 

Due to the symmetry breaking of the electric fields it is no longer possible to separate the relative motion of the particles from the motion of the center of mass, and therefore the relative and convective states are mixed. In this paper we solve this intricate problem in the case when the BEP has a zero total momentum without any restrictions on the relative magnitude of the tangential and normal components of the electric field. As a result, we came to the conclusion that the in-plane component plays a key role in the formation of the BEPs, and the presence of a normal component leads to the radical reconstruction of the BEPs.

Specific calculations are performed for a model system that consists of an atomically thin layer of material with a strong Rashba SOI separated by a spacer from a charged metallic gate, as shown in Fig.~\ref{fig1}. The presence of the gate affects both the spatial configuration and the magnitude of the Coulomb field of the interacting electrons. In addition, the external voltage applied to the gate creates one-particle Rashba SOI\@. We explore how these factors govern the BEP formation, their spectrum and spin structure. 

The normal field lifts the degeneracy of the relative BEPs to produce two kinds of BEPs having very different properties. Our most interesting finding is that there appears a \emph{robust} BEP that remains unchanged with the variation of the gate potential. At large enough gate voltage of any sign, the robust BEP gets into the continuum of the band states, where it remains localized. On the contrary, the BEP of the other kind is tunable by the gate voltage. The positive voltage applied to the gate increases its binding energy, whereas the negative voltage moves the energy level of the \emph{tunable} BEP to the continuum, where it decays.

\section{The model}

Consider two electrons at positions $\mathbf{r}_{i}$ in the 2D system. In the two-particle basis $\{\lvert \uparrow \uparrow  \rangle, \lvert \uparrow \downarrow  \rangle, \lvert \downarrow \uparrow  \rangle, \lvert \downarrow \downarrow  \rangle \}$ the system wave-function represents a Pauli spinor of the fourth rank, $\Psi(\mathbf{r}_{1},\mathbf{r}_{2}) = {\left(\Psi_{\uparrow \uparrow},\Psi_{\uparrow \downarrow},\Psi_{\downarrow \uparrow},\Psi_{\downarrow \downarrow}\right)}^{\intercal}$. Introduce the relative electron position $\mathbf{r} = \mathbf{r}_1 - \mathbf{r}_2$, the center-of-mass position $\mathbf{R} = (\mathbf{r}_{1} + \mathbf{r}_{2})/2$, and the corresponding momenta $\mathbf{p} =  -i \hbar\nabla_{\mathbf{r}}$ and $\mathbf{P} =  -i \hbar\nabla_{\mathbf{R}}$. The PSOI Hamiltonian, built as the Kronecker sum of the terms in Eq.~\eqref{pse}, is equal to
\begin{widetext}
	\begin{equation}
		\label{PSOI}
			H_{\mathrm{PSOI}} = \frac{\alpha}{2 \hbar}
				\left(
					\begin{matrix}
						\frac{4 E_{\tau}(r)}{r} {(\mathbf{r} \times \mathbf{p})}_{z} && -\xi_{+} + \Xi_{+} && \xi_{+} + \Xi_{+} && 0 \\
						-\xi_{-} + \Xi_{-} && \frac{2 E_{\tau}(r)}{r} {(\mathbf{r} \times \mathbf{P})}_{z} && 0 && \xi_{+} + \Xi_{+} \\
						\xi_{-} + \Xi_{-} && 0 && -\frac{2 E_{\tau}(r)}{r} {(\mathbf{r} \times \mathbf{P})}_{z} && -\xi_{+} + \Xi_{+} \\
						0 && \xi_{-} + \Xi_{-} && -\xi_{-} + \Xi_{-} && -\frac{4 E_{\tau}(r)}{r} {(\mathbf{r} \times \mathbf{p})}_{z}
					\end{matrix}
				\right)\,,
	\end{equation}
\end{widetext}
with $\xi_{\pm} = {[F(r),\gamma_{\pm}]}_{+}$ and $\Xi_{\pm} = F(r) \Gamma_{\pm}$. The normal field $F(r) \equiv E_{n}(r)$ is given by Eq.~\eqref{effield} for the particular geometry considered in Fig.~\ref{fig1}. Then $\Gamma_{\pm} = P_{y} \pm i P_{x}$, and
\begin{equation}
		\gamma_{\pm} = p_y \pm i p_{x} = \hbar e^{\mp i \phi} \left(\pm \partial_{r} - \frac{i}{r}\partial_{\phi} \right)\,.
\end{equation}
The anti-commutator ${[\hat{A}, \hat{B}]}_{+} = \hat{A}\hat{B} + \hat{B}\hat{A}$ is introduced to maintain the hermiticity of the Hamiltonian when projecting Eq.~\eqref{pse} to the 2D subspace.

A single-particle SOI contribution can be included in Eq.~\eqref{PSOI} by adding the field 
\begin{equation}
\label{gatef}
	F_{g} = 4 \pi n_{g} + E_n(0)\,,
\end{equation}
produced by the gate surface charge density $n_{g}$ and the electron own image, to the normal field  $F(r)$, so that the total normal field becomes $F(r) = E_{n}(r) + F_{g}$.

Below we restrict ourselves to a particular case of $P = 0$, when the BEPs are essentially the relative ones.

The equation of motion for the two-body wave-function follows from the full Hamiltonian 
\begin{equation}
\label{fullH}
	H = H_\mathrm{PSOI} + V + T\,,
\end{equation}
which in addition to the PSOI of Eq.~\eqref{PSOI} contains diagonal contributions coming from the electron-electron (e-e) repulsion $V(r)$ of Eq.~\eqref{pot2} and the kinetic energy $T$. In what follows it is convenient to introduce the shift  $e F_{g} a$ in the energy and the potential $V$. This eliminates the trivial effect of the gate potential and allows us to consider only the effect of the normal electric field. For simplicity, we consider here  a minimal model with quadratic band dispersion.

In the absence of the gate, when $F(r) \equiv 0$, the relative BEPs represent degenerate pairs of triplet states with the spin orientation tied to the angular momentum direction~\cite{PhysRevB.98.115137,GINDIKIN2019187}. The lowest-lying states, corresponding to the minimum possible angular momentum $l = \pm 1$, are
\begin{equation}
\label{deg1}
	\Psi_{-}(\mathbf{r}) = {\left(u(r) e^{- i \phi},0,0,0\right)}^{\intercal}
\end{equation}
and 
\begin{equation}
\label{deg2}
	\Psi_{+}(\mathbf{r}) = {\left(0,0,0,u(r) e^{i \phi}\right)}^{\intercal}\,.
\end{equation}
The radial wave-function $u(r)$ is determined from the Schrödinger equation
\begin{equation}
\label{rad}
		\left[ T_{1} +V(r) - 2 \alpha \frac{E_{\tau}(r)}{r} \right] u(r) = \varepsilon_{0} u(r) \,,
\end{equation}
where $T_{l}$ stands for the kinetic energy including the centrifugal potential,
\begin{equation}
	T_{l} = -\frac{\hbar^2}{m} \left( \frac{d^{2}}{d r^{2}} + \frac{1}{r} \frac{d}{d r} - \frac{l^2}{r^2} \right), \quad l = 0, \pm 1, \ldots \,.
\end{equation}

The binding potential produced by PSOI is the last term on the left hand side of Eq.~\eqref{rad}. Taking into account the short-range asymptotics of the in-plane field $E_{\tau}$ given by Eq.~\eqref{astau}, we see that the BEPs are formed by the singular attractive potential  $\propto -\dfrac{\alpha}{\chi r^{2}}$, with $\chi$ being the 2D susceptibility of the layer. This overcomes the centrifugal barrier for sufficiently large $\alpha$, let alone much weaker Rytova-Keldysh repulsion $\propto \log \frac{r}{2 \pi \chi}$. 

The $-1/r^{2}$ potential leads to a fall to the center~\cite{landau1958course}, unless a short-range cut-off is introduced. Regularization of the binding potential can be caused by mechanisms such as the Zitterbewegung of electrons in crystalline solids or natural cutting-off due to averaging the three-dimensional quantities across the layer thickness. We regularize the potential by imposing a zero boundary condition for the wave-function at the cut-off length of the order of the the layer thickness~\cite{PhysRevB.98.115137,GINDIKIN2019187}. Then binding energy can be estimated as
\begin{equation}
\label{spec0}
	|\tilde{\varepsilon}_{0}| = \frac{x_{1}^{2} (\lambda)}{{(d/a_{B})}^{2}} \,,
\end{equation}
where $x_{1} (\lambda)$ is the first (largest) zero of the Macdonald function $K_{i \lambda}(x)$, and the amplitude of the attraction is defined as
\begin{equation}
	\lambda = \sqrt{\frac{4 \tilde{\alpha}}{d/a_{B}} - 1} \,.
\end{equation}
Here we introduced a convenient dimensionless SOI constant $\tilde{\alpha} = \alpha/e a_{B}^{2}$, with the Bohr radius in the material $a_{B} = \epsilon \hbar^2/m e^2$. The BEPs appear as soon as $\tilde{\alpha} > d/4 a_{B}$, which is attainable in materials like $\mathrm{Bi_2 Se_3}$~\cite{PhysRevLett.107.096802}, $\mathrm{BiTeI}$~\cite{ishizaka2011giant,F_l_p_2018} or $\mathrm{BiSb}$ monolayers~\cite{PhysRevB.95.165444}, for which $\tilde{\alpha}$ is of the order of unity~\cite{manchon2015new}.
From now on, the energy with a tilde is given in $2 Ry$ units, with the Rydberg constant in the material being $Ry = \hbar^2/2 m a_B^2$. Eq.~\eqref{spec0} gives $|\varepsilon_{0}|$ on the level of tens of Rydberg.

\section{Robust BEP}
The normal field $F(r)$ that appears in the presence of the gate lifts the degeneracy. In the lowest order of degenerate perturbation theory, a perturbation does so by mixing the states with certain weights defined by its matrix elements~\cite{landau1958course}. However, the corresponding matrix elements calculated with the states of Eqs.~\eqref{deg1}--\eqref{deg2} are all zero. Consequently, a higher order approximation should be used which, generally speaking, involves the scattering states of the Hamiltonian of Eq.~\eqref{fullH} in the perturbation expansion. Fortunately, this tedious procedure can be avoided by checking that the state
\begin{equation}
\label{prot}
		\Psi(\mathbf{r}) = {\left(u(r)e^{- i \phi},0,0, -u(r) e^{i \phi}\right)}^{\intercal}
\end{equation}
with $u(r)$ given by Eq.~\eqref{rad} provides the exact solution of the full Hamiltonian~\eqref{fullH} for the arbitrary magnitude of the normal field $F(r)$. This antisymmetric combination of the unperturbed solutions of Eqs.~\eqref{deg1}--\eqref{deg2} obviously does not include any scattering states, which would depend on the normal field $F(r)$.

It follows from Eq.~\eqref{rad} that neither the radial wave-function $u(r)$, nor the energy $\varepsilon_{0}$ depend on the normal field $F(r)$. Therefore the bound state of Eq.~\eqref{prot} is robust with its orbital and spin structure unaffected by the normal electric field applied to the system. The result is not specific to a particular sandwich geometry considered here and holds for any profile of $F(r)$ provided that i) there is no external  field parallel to the layer and ii) the electron pair has a zero total momentum.

Note that the binding energy is measured from the bottom of the conduction band $\varepsilon_{c}$, which in the presence of the SOI is shifted as soon as $F_{g} \ne 0$. For a pair of electrons, its position is given by 
\begin{equation}
	\tilde{\varepsilon}_{c} = - \frac{1}{4} \mathcal{F}_{g}^2 \,,
\end{equation}
where $\mathcal{F}_{g}$ stands for the gate field $F_{g}$ of Eq.~\eqref{gatef} normalized according to $\mathcal{F}_{g} = \tilde{\alpha} F_{g}/F_{0}$, with $F_{0} = e/2 \epsilon a_{B}^{2}$. In 1D quantum wires the BEPs always lie below $\varepsilon_{c}$~\cite{2018arXiv180410826G}. This is not, generally speaking, the case in a 2D system. Increasing $F_{g}$ lowers $\varepsilon_{c}$, keeping $\varepsilon_{0}$ intact, so eventually the energy level $\varepsilon_{0}$ gets into the conduction band. According to Eq.~\eqref{rad}, the robust BEP remains localized even at $\varepsilon_{0} > \varepsilon_{c}$. In other words, there appears a discrete energy level in the continuum that does not mix with the band states.

\section{Tunable BEP}
Contrary to Eq.~\eqref{prot}, the symmetric combination of Eqs.~\eqref{deg1}--\eqref{deg2} is not a solution at $F \ne 0$. All four spinor components do arise in the exact solution, which reads as
\begin{equation}
\label{BS}
	\Psi(\mathbf{r}) = {\left(u(r)e^{- i \phi}, v(r), -v(r), u(r) e^{i \phi}\right)}^{\intercal} \,.
\end{equation}
This form ensures the anti-symmetry of $\Psi$ with respect to the permutation of electrons. From the point of view of the perturbation theory, Eq.~\eqref{BS} includes the contribution from the scattering states of Eq.~\eqref{fullH}, which makes it sensitive to the normal field $F(r)$.

The radial wave functions satisfy the system of equations
\begin{numcases}{}
\label{bsys}
	\left[T_{1} +V(r) - 2 \alpha \frac{E_{\tau}(r)}{r} \right] u -\alpha \left[2 F(r) \frac{d}{d r} + F'(r) \right] v = \varepsilon u & \nonumber \\[1ex]
	\left[T_{0} +V(r) \right] v +\alpha \left[2 F(r) \left(\frac{1}{r} + \frac{d}{d r} \right) + F'(r) \right] u = \varepsilon v\,, &
\end{numcases}
which should be solved with zero boundary conditions at $r = d$ and at infinity. We are mostly interested in the dependence of the binding energy on the normal electric field. 

The analytical treatment of this problem is expounded in Appendix~\ref{msasec}. Here we solve the Eq.~\eqref{bsys} numerically with the exact interaction potential and field of Eq.~\eqref{pot2} and Eq.~\eqref{effield}. To give an estimate of the binding energy of the tunable state, consider the system based on a $\mathrm{Bi_2 Se_3}$, for which $\alpha \approx \SI{1300}{e {\angstrom}^2}$~\cite{manchon2015new}, $a_B \approx \SI{52}{\angstrom}$ and hence $\tilde{\alpha} \approx 0.47$. For a reasonable value of the electric field $F_g = \SI{e5}{V/cm}$, the layer thickness of  $d =  \SI{28.7}{\angstrom}$, corresponding to three quintuple layers of $\mathrm{Bi_2 Se_3}$, and the distance to the gate $a = 2d$, we obtain $|\varepsilon| = \SI{40}{meV}$.

\begin{figure}[t]
	\centering
	\includegraphics[width=0.9\linewidth]{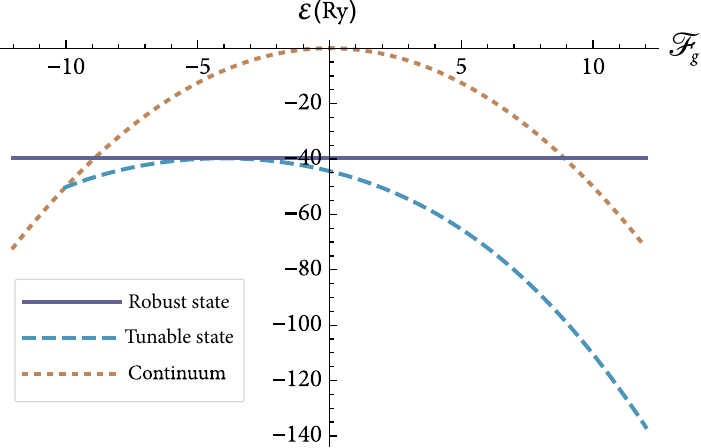}
		\caption{The dependence of the binding energy of the BEPs, as well as the bottom of the conduction band, on the normalized gate field.
	\label{fig2}}
\end{figure}

In Fig.~\ref{fig2} we plot the energy levels of the robust BEP of Eq.~\eqref{prot} and the tunable BEP of Eq.~\eqref{BS} vs.\ the normalized field of the gate for the model system with $\tilde{\alpha}=1$, $\chi = 0.4 a_{B}$, $d = 0.25 a_{B}$ and $a = a_{B}$. Additionally, the position of the bottom of the conduction band is shown. At large negative voltage applied to the gate, the tunable BEP gets into the continuum where it decays, whereas the positive voltage facilitates the pairing by increasing the binding energy. The binding energy of the robust BEP, measured from $\varepsilon_{c}$, is decreasing when the voltage is applied to the gate, so at large gate voltage of any polarity the energy level crosses the continuum boundary, but the robust BEP remains bound and localized even in the continuum.

\section{Conclusion}
We studied the BEPs formed by the PSOI in the most realistic and practically important situation of a 2D system with a gate, when the PSOI is created by a Coulomb field of interacting electrons having both in-plane and normal components. We focus on the effects due to the interplay of these components for a particular case of a BEP with zero total momentum. 

We have found that the normal field lifts the degeneracy of the pair of bound states created by the in-plane field. One of the resulting BEPs that has a higher binding energy is tunable by the gate voltage. Its binding energy is significantly increased by the positive gate voltage, and its spin and orbital structure continuously transforms when changing $\mathcal{F}_g$.

In contrast, the second state demonstrates totally unexpected behavior. Its spin and orbital structure does not depend on the gate voltage, and its energy varies exactly as the potential induced by the gate at the layer. The binding energy measured from the conduction band bottom decreases with the gate voltage, so at large $\mathcal{F}_g$ the energy level gets into the continuum of the band states. It is interesting that this state remains bound and localized even when it is in the continuum. The fact that the BEP is so stable with respect to the normal electric field evidences that the in-plane electric field of the Coulomb interaction plays a key role in the electron pairing in competition with the normal component.

The behavior of electrons in the presence of the PSOI in a many-electron system needs further serious study. One can expect that because of the unusual form of the pair interaction, various scenarios are possible such as the formation of electronic complexes, spontaneous symmetry breaking, and of course the formation of superconducting phases.

\begin{acknowledgements}
	This work was carried out in the framework of the state task for IRE RAS and partially was supported by the Presidium of the Russian Academy of Sciences, Program No13 ``Fundamentals of high technology and the use of features of nanostructures in the natural science'', and by the Russian Foundation for Basic Research, the project No 20--02--0126.
\end{acknowledgements}

\appendix
\section{Electrostatics of the 2D gated layer}
In this section we derive the e-e interaction potential in the gated layer, as well as the electric field acting on the electrons, which defines the PSOI Hamiltonian.

Consider a charge $-e$ in a 2D layer placed at the origin $(\mathbf{r} = 0, z = 0)$, where $\mathbf{r}$ stands for the in-plane position. Let us find the potential $\varphi (\mathbf{r})$ the charge creates at certain point $\mathbf{r}$ at the layer.

The layer is assumed to be extremely thin. The dielectric properties of the 2D layer are properly described by a 2D susceptibility $\chi$, rather than a dielectric constant~\cite{PhysRevB.84.085406,doi:10.1063/1.5052179}. The perpendicular susceptibility $\chi_{n}$ is very small in comparison with $\chi$ for atomically thin layers~\cite{PhysRevB.88.195410}, and will be neglected below.

The susceptibility $\chi$ relates the 2D polarization vector $\mathbf{P}_{\tau}$, defined as the dipole moment of a surface element, with the in-plane field $\mathbf{E}_{\tau}$ via 
\begin{equation}
	\mathbf{P}_{\tau} = \chi \delta(z) \mathbf{E}_{\tau} = - \chi \delta(z) \nabla_{r} \varphi \,,
\end{equation}
so the polarization charge density is
\begin{equation}
\label{polden}
	\rho = - \nabla \cdot \mathbf{P}_{\tau} = \chi \delta(z) \Delta_{r} \varphi \,.
\end{equation}
The Poisson equation is then
\begin{equation}
	\Delta_{r,z} \varphi = 4 \pi e \delta(\mathbf{r}) \delta(z) - 4 \pi \chi \delta(z) \Delta_{r} \varphi \,,
\end{equation}
which, after a 2D Fourier-transform, becomes
\begin{equation}
	\varphi_{zz} - k^2 \varphi = 4 \pi e \delta(z) + 4 \pi \chi k^2 \delta(z) \varphi \,.
\end{equation}

The solution, corresponding to the zero boundary condition at the gate, is 
\begin{numcases}{\varphi (k,z) = - \frac{4 \pi e }{k}}
\label{pot1}
	\frac{\sinh k(z + a)}{e^{k a} + 4 \pi \chi k \sinh (ka)}, & $-a \le z \le 0$\,, \nonumber \\
	\frac{e^{- k z} }{1 + 4 \pi \chi k + \coth (k a)}, & $z \ge 0 $\,.
\end{numcases}

The e-e interaction potential is given by
\begin{equation}
\label{pot2}
	V(r) = -e \varphi (r,z) \big \vert_{z=0} = 2 e^2 \int_{0}^{\infty} \frac{J_{0}(k r)\, dk}{1 + 4 \pi \chi k + \coth (k a) }\,,
\end{equation}
with $J_{0}$ being the Bessel function of the first kind. At $r \ll a$ this turns into the Rytova-Keldysh potential~\cite{rytova,keldysh1979coulomb}.

The effective electric field $\mathbf{E}(\mathbf{r}) \equiv (\mathbf{E}_{\tau},E_{n})$ acting on electrons depends on the microscopic model of the 2D layer. To be specific, we suppose that the moving electrons are located symmetrically in the close vicinity of the layer. In this case,
\begin{equation}
\label{effield}
	\begin{split}
		E_{n}(r) &= \frac{1}{2} \left( -\frac{\partial \varphi}{\partial z} \bigg \vert_{z = -0} - \frac{\partial \varphi}{\partial z} \bigg \vert_{z = +0} \right)\\
		&= e \int_{0}^{\infty} \frac{J_{0}(kr) e^{-ka} k\, dk}{e^{k a} + 4 \pi \chi k \sinh (ka)}\,.
	\end{split}
\end{equation}
The in-plane field is determined from Eq.~\eqref{pot2} via $\mathbf{E}_{\tau}(\mathbf{r}) = \frac{1}{e} \nabla_{r} V$. 

The 2D susceptibility, measured in cm, can be estimated as $\chi \approx \epsilon d/4 \pi$, where $\epsilon$ is the in-plane component of the dielectric tensor of the bulk material, and $d$ is the layer thickness~\cite{PhysRevB.88.045318}. In materials with a strong Rashba effect, $\epsilon$ is typically large, somewhere in between $10 \ldots 100$~\cite{manchon2015new}. We are interested in a particular case of $a < 4 \pi \chi$ to fully involve the image charges induced on the gate. At $r \ll a$, we have
\begin{equation}
\label{nas}
	E_{n}(r) \approx \frac{e}{2 \pi \chi} \frac{1}{\sqrt{r^2 + 4 a^2}}
\end{equation}
and 
\begin{equation}
\label{astau}
	\mathbf{E}_{\tau}(\mathbf{r}) \approx - \frac{e}{2 \pi \chi r} \frac{\mathbf{r}}{r}\,.
\end{equation}

\section{Multiple-scale Analysis}
\label{msasec}

The analytic approach to Eq.~\eqref{bsys} is based on the fact that the bound states in the singular attractive potential $\propto -1/x^2$ are formed on the scale of the short-range cut-off $d$, which is assumed much smaller that the distance to the gate. On this scale, the normal field $E_{n}(r)$ of Eq.~\eqref{nas} can be considered constant $E_{n}(r) \approx E_{n}(0)$, so the total normal field becomes homogeneous,
\begin{equation}
\label{horshift}
	F = E_{n}(0) + F_{g} =  4\pi n_{g} + 2E_{n}(0)\,.
\end{equation}
Also, the weaker $\sim \log r$ repulsive potential of Eq.~\eqref{pot2} can be neglected there. Let us normalize $r$ and $d$ to $a_B$ and introduce the dimensionless total normal field $\mathcal{F} = \tilde{\alpha} F/F_{0}$. Then the system takes the form 
\begin{equation}
\label{nsys}
	\begin{dcases}
		\begin{aligned}
			-u'' - \frac{1}{r} u' - \frac{\lambda^2}{r^2} u - \mathcal{F} v' &= \tilde{\varepsilon} u \\
			-v'' - \frac{1}{r} v' + \mathcal{F}\left(u' + \frac{u}{r}\right) &= \tilde{\varepsilon} v \,.
		\end{aligned}
	\end{dcases}
\end{equation}

The system can be easily solved at $\mathcal{F} = 0$. However, the direct perturbative expansion of the solution in powers of $\mathcal{F}$ has a very limited range of applicability, because the solution reveals an oscillatory behavior in $\mathcal{F}$, so that secular terms appear in the expansion. Instead of the conventional perturbation expansion, we use the multiple-scale analysis (MSA)~\cite{PhysRevD.54.7710,bender1999advanced}.

\begin{figure}[htb]
	\includegraphics[width=0.9\linewidth]{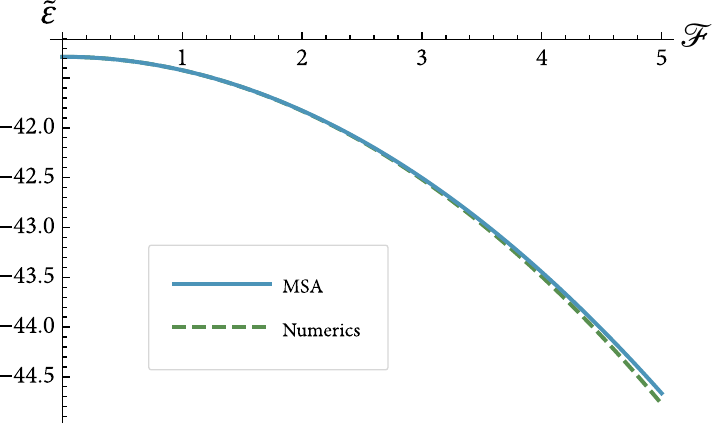}
		\caption{The dependence of the binding energy on the total normal field. Here $\tilde{\alpha}=1$ and $d = 0.25 a_{B}$.
	\label{fig3}}
\end{figure}

\begin{figure}[htb]
	\includegraphics[width=0.9\linewidth]{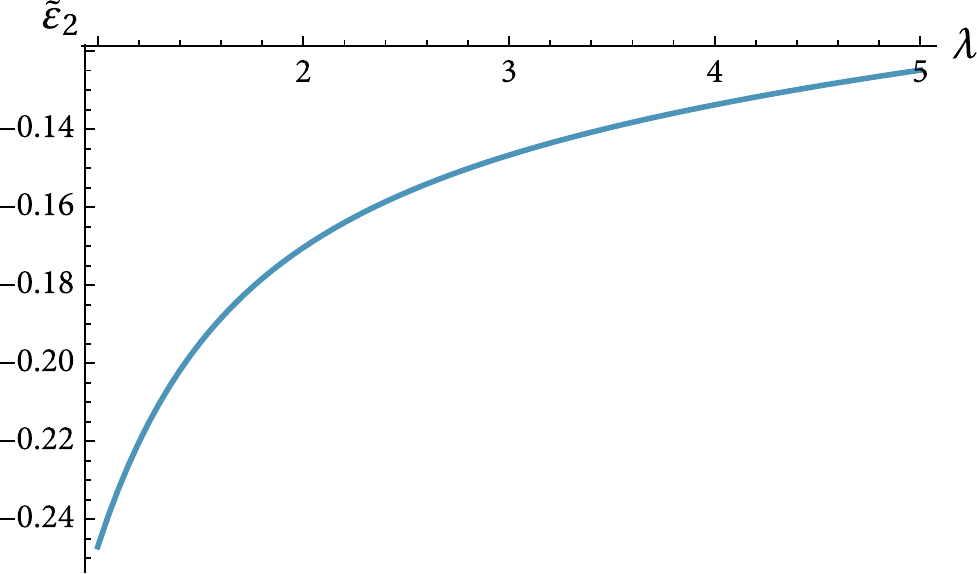}
		\caption{The dependence of the second order energy correction on the amplitude of the attractive potential.
	\label{fig4}}
\end{figure}

Taking into account that the expansion of $u$ should start with $\mathcal{F}^{0}$ and that upon the change $\mathcal{F} \to -\mathcal{F}$ the pair of functions $(u,-v)$ still gives the solution of the system, we conclude that $u$ is expanded in even powers of $\mathcal{F}$, and $v$ --- in odd powers. Introduce along with $r$ the long-range scale $\xi_{2} = \mathcal{F}^{2} f_{2}(r)$, where the unknown function $f_{2}(r)$ is yet to be determined. In MSA, the scales $r$ and $\xi_{2}$ are treated as independent variables. Restricting the expansion to the second order in $\mathcal{F}$, we have
\begin{equation}
\label{msaexp}
	\begin{split}
		u(r) &= u_{0}(r,\xi_{2}) + \mathcal{F}^{2}u_{2}(r,\xi_{2}) + \mathcal{O}(\mathcal{F}^{4})\,, \\
		v(r) &= \mathcal{F} v_{1}(r,\xi_{2}) + \mathcal{O}(\mathcal{F}^{3})\,,\\
		\tilde{\varepsilon} &= \tilde{\varepsilon}_{0} + \mathcal{F}^{2}\tilde{\varepsilon}_{2} + \mathcal{O}(\mathcal{F}^{4})\,. 
	\end{split}
\end{equation}
The derivatives in Eq.~\eqref{nsys} are
\begin{align}
	&\frac{du}{dr} = \frac{\partial u_{0}}{\partial r} + \mathcal{F}^2\left( f_{2}'\frac{\partial u_{0}}{\partial \xi_{2}} + \frac{\partial u_{2}}{\partial r} \right) + \mathcal{O}(\mathcal{F}^{4})\,,\notag \\
	&\frac{d^{2} u}{d r^{2}} = \frac{\partial^{2} u_{0}}{\partial r^{2}} + \mathcal{F}^2\left( f_{2}''\frac{\partial u_{0}}{\partial \xi_{2}} + 2f_{2}'\frac{\partial^{2} u_{0}}{\partial r \partial \xi_{2}} + \frac{\partial^{2} u_{2}}{\partial r^{2}} \right) + \mathcal{O}(\mathcal{F}^{4})\,,\notag \\
	&\frac{dv}{dr} = \mathcal{F} \frac{\partial v_{1}}{\partial r} +  \mathcal{O}(\mathcal{F}^{3})\,,\\
	&\frac{d^{2} v}{d r^{2}} = \mathcal{F} \frac{\partial^{2} v_{1}}{\partial r^{2}} + \mathcal{O}(\mathcal{F}^{3})\,.\notag 
\end{align}
Then, in each order of $\mathcal{F}$, we have:

$\mathcal{F}^{0}$) The equation for $u_{0}$ is
\begin{equation}
	-\partial_{r}^{2} u_{0} - \frac{1}{r}\partial_{r} u_{0} -\frac{\lambda^2}{r^2} u_{0} = - \kappa^{2} u_{0}\,,
\end{equation}
with $\kappa = \sqrt{|\tilde{\varepsilon}_{0}|}$. Its solution is 
\begin{equation}
\label{expl}
	u_{0} = A(\xi_{2}) K_{i \lambda} (\kappa r)\,,
\end{equation}
where $A(\xi_{2})$ is an unknown function to be determined. The spectrum is defined from the boundary condition of $u_{0}(d) = 0$, which yields $\kappa d = x_1(\lambda)$, where $x_1 (\lambda)$ is the first zero of $K_{i \lambda}(x)$. This leads to Eq.~\eqref{spec0} for $\tilde{\varepsilon}_{0}$.

$\mathcal{F}^{1}$) The equation for $v_{1}$ is
\begin{equation}
	-\partial_{r}^{2} v_{1} - \frac{1}{r}\partial_{r} v_{1} + \partial_{r} u_{0} + \frac{u_{0}}{r} = - \kappa^{2} v_{1}\,.
\end{equation}
We require that its solution $v_{1}(r,\xi_{2}) =  A(\xi_{2}) w(r)$ be defined by the same function of $\xi_{2}$ as $u_{0}$. The solution that satisfies the zero boundary condition at $r=d$ is
\begin{widetext}
	\begin{equation}
		w(r) = K_{0}^{-1}(\kappa d) \left[ \left( K_{0}(\kappa r)I_{0}(\kappa d) - K_{0}(\kappa d)I_{0}(\kappa r) \right) \int_{r}^{\infty} K_0 (\kappa \eta) g(\eta) \, d\eta + K_{0}(\kappa r) \int_{d}^{r} \left( K_{0}(\kappa \eta)I_{0}(\kappa d) - K_{0}(\kappa d)I_{0}(\kappa \eta) \right) g(\eta) \, d\eta \right]\,,
	\end{equation}
\end{widetext}
where $I_{0}(\kappa r)$ is a modified Bessel function, and
\begin{equation}
	g(\eta) = \kappa \eta K_{i \lambda}'(\kappa \eta) + K_{i \lambda}(\kappa \eta)\,.
\end{equation}

$\mathcal{F}^{2}$) Only now MSA seriously comes into play. The equation for $u_{2}$ is
\begin{equation}
	\begin{split}
		&-\partial_{r}^{2} u_{2} - \frac{1}{r}\partial_{r} u_{2} -\frac{\lambda^2}{r^2} u_{2} - \tilde{\varepsilon}_{0} u_{2} =\\
		&f_{2}''\frac{\partial u_{0}}{\partial \xi_{2}} + 2f_{2}'\frac{\partial^{2} u_{0}}{\partial r \partial \xi_{2}}+\frac{1}{r} f_{2}'\frac{\partial u_{0}}{\partial \xi_{2}} + \tilde{\varepsilon}_{2} u_{0} + A(\xi_{2}) \frac{\partial w}{\partial r}\,.
	\end{split}
\end{equation}
The essence of the method is to require that the right hand side of the equation be equal to zero to prevent the secular terms from appearing in $u_{2}$~\cite{PhysRevD.54.7710}. Substitute here Eq.~\eqref{expl} for $u_{0}$ to obtain
\begin{equation}
	\frac{A'(\xi_{2})}{A(\xi_{2})} = - \frac{\tilde{\varepsilon}_{2} K_{i \lambda}(\kappa r) + \frac{\partial w}{\partial r}}{f_{2}''K_{i \lambda}(\kappa r) + f_{2}'\left( 2 \kappa K_{i \lambda}'(\kappa r) + \frac{1}{r} K_{i \lambda}(\kappa r)\right)}\,.
\end{equation}
The variables $r$ and $\xi_{2}$ in this equation are separated, hence both sides are equal to the same constant, which without loss of generality can be taken as $1$, because it is scaled out in the end. Then
\begin{equation}
	f_{2}'(r) = \frac{1}{r K_{i \lambda}^{2}(\kappa r)} \int_{d}^{r} \eta K_{i \lambda}^{2}(\kappa \eta)\left( \tilde{\varepsilon}_{2} + \frac{1}{K_{i \lambda}(\kappa \eta)}\frac{\partial w}{\partial \eta} \right)\, d \eta \,.
\end{equation}
To suppress the secular growth of $f_{2}'(r)$ at $r \to \infty$, there should be
\begin{equation}
	\tilde{\varepsilon}_{2} = - \frac{\int_{d}^{\infty} \eta K_{i \lambda}(\kappa \eta)\frac{\partial w}{\partial \eta}\, d \eta}{\int_{d}^{\infty} \eta K_{i \lambda}^2(\kappa \eta)\, d \eta}\,.
\end{equation}
This is our final result. The dependence of the binding energy $\tilde{\varepsilon} = \tilde{\varepsilon}_{0} + \mathcal{F}^2 \tilde{\varepsilon}_{2}$ on the total normal field $\mathcal{F}$ is illustrated in Fig.~\ref{fig3}. For comparison, the result of the numerical solution of Eq.~\eqref{nsys} is also shown. A good agreement is seen up to $\mathcal{F} = 5$, with this range getting wider as $\lambda$ increases.

The dependence of $\tilde{\varepsilon}_{2}(\lambda)$ is shown in Fig.~\ref{fig4}. The larger the $\lambda$, the less steep the dependence of the binding energy on the normal field. In other words, a deep state barely feels the normal field. 

At large $\lambda$, we have $|\tilde{\varepsilon}_{2}| < 0.25$, which means that the curve $\tilde{\varepsilon}_{c}(\mathcal{F}_{g}) = - \mathcal{F}_{g}^2/4$ that shows the position of the bottom of the conduction band goes steeper than the energy level of the bound state $\tilde{\varepsilon}(\mathcal{F}) = \tilde{\varepsilon}_{0} + \mathcal{F}^2 \tilde{\varepsilon}_{2}$. Besides, the maxima of the two parabolas are shifted against each other in accordance with Eq.~\eqref{horshift}. Hence the curves intersect at some point, where the energy level of the bound state of Eq.~\eqref{BS} gets into the conduction band.

Once in a conduction band, the bound state decays. This is clearly seen from the asymptotic form of the solution of Eq.~\eqref{bsys} at $r \to \infty$, which reads as $u(r) \sim K_{1}(kr)$ and $v(r) \sim i K_{0}(kr)$, with
\begin{equation}
\label{asimptot}
	k = i \sqrt{|\tilde{\varepsilon}_{c}|} + \sqrt{\tilde{\varepsilon}_{c} - \tilde{\varepsilon}}\,.
\end{equation}
If $\tilde{\varepsilon} > \tilde{\varepsilon}_{c}$, $k$ is pure imaginary, which means that the state becomes de-localized.

The numerical calculations of the main section confirm the MSA results.

\bibliography{paper}

\end{document}